\documentclass[12pt]{iopart}

\usepackage{iopams}
\usepackage{graphicx}
\usepackage{bm}         

\newcommand{\cg}{\textnormal{\textsl{g}}}
\newcommand{\cD}{\mathcal{D}}
\newcommand{\cT}{\mathcal{T}}
\newcommand{\cS}{\mathbb{S}}
\newcommand{\cG}{\check{G}}
\newcommand{\cU}{\check{U}}
\newcommand{\cz}{\tilde{z}}
\newcommand{\cW}{\hat{W}}
\newcommand{\llangle}{\langle\langle}
\newcommand{\rrangle}{\rangle\rangle}
\newcommand{\floor}[1]{\lfloor #1 \rfloor}
\newcommand{\ceil}[1]{\lceil #1 \rceil }
\newcommand{\csch}{\textnormal{csch}}
\usepackage{color} 			

\begin{document}
    

\title{Non-classical current noise and light emission of an ac-driven tunnel junction}

\author{Hongxin Zhan, Gianluca Rastelli, Wolfgang Belzig}
\address{Fachbereich Physik, Universit\"{a}t Konstanz, D-78457 Konstanz, Germany}

\date{\today}

%
\begin{abstract}
The non-symmetrized current noise is crucial for the analysis of light emission in nanojunctions. The latter represent non-classical photon emitters whose description requires a full quantum approach. It was found experimentally that light emission can occur with a photon energy exceeding the applied dc voltage, which intuitively should be forbidden due to the Pauli principle. This overbias light emission cannot be described by the single-electron physics, but can be explained by two-electron or even three-electron processes, correlated by a local resonant mode in analogy to the well-known dynamical Coulomb blockade (DCB). Here, we obtain the non-symmetrized noise for junctions driven by an arbitrarily shaped periodic voltage. We find that when the junction is driven, the overbias light emission exhibits intriguingly different features compared to the dc case. In addition to kinks at multiples of the bias voltage, side kinks appear at integer multiples of the ac driving frequency. Our work generalizes the DCB theory of light emission to driven tunnel junctions and opens the avenue for engineered quantum light sources, which can be tuned purely by applied voltages.
\end{abstract}

\section{Introduction}
The study of the quantum nature of shot noise \cite{Buttiker_shot_noise_review_2000} has progressed constantly since the first prediction of shot noise by Schottky \cite{Schottky_shot_noise_1918} in 1918.  Shot noise, as a consequence of the discreteness of the electric charge carrier, reveals information which the standard electrical current measurement cannot give. It has subsequently played a major role in unraveling fractional charges \cite{dePicciotto1997,Saminadayar1997} and multiple charges in superconducting transport \cite{Khlus:1987,Muzy:94,Fauchere:1998wz,Cuevas:1999um,Jehl_detection_of_doubled_2000,Cron:2001vc,Cuevas:2003dta,Johansson:2003ff,Lefloch:2003fp,Cuevas:2004ec}. More subtle coherence effects can also be detected through noise measurements in Andreev interferometers  \cite{Belzig:2001kw,Reulet:2003dg}. 
Furthermore, noise has for example been used to characterize entanglement via the violation of Bell-type inequalities  \cite{Chtchelkatchev_Bell_inequality_2002,Zhan:2019bo,LESOVIK:2001tr}, to probe many-body states of ultracold atoms \cite{Altman_probe_many_body_2004}, to reveal the relativistic quantum dynamics in graphene \cite{Tworzydlo_sub_poissonian_shot_noise_2006,Hammer:2013gb}, to observe Majorana bound states in p-wave superconductors \cite{Bolech_observing_majorana_2007} and in many more applications. 

Owing to the developments of theory and experimental techniques in quantum transport, the quantum shot noise has been also intensively investigated over the recent decades \cite{Lesovik_on_the_1997,Jehl_detection_of_doubled_2000,Gavish_detection_of_quantum_2000,Aguado_double_quantum_2000,Oberholzer_crossover_between_2002,Deblock_detection_of_quantum_noise_2003,Hammer_quantum_noise_2011,Stadler_finite_frequency_2018}.
Classically, the noise spectrum $S(\omega)=\int dt e^{i\omega t} \langle\Delta I(t)\Delta I(0)\rangle$ of a conductor is symmetrized in frequency $S(\omega)=S(-\omega)$. However, for a quantum conductor the current operators $\hat{I}$ do not commute with each other at different times, leading to a spectrum that is not symmetric in frequency i.e. $S(\omega)\neq S(-\omega)$ which manifests the quantum nature of noise. Indeed light emission is related to the non-symmetrized noise and the measured operator order depends e.g. on the memory of the detector \cite{Bednorz:2013bj,Bulte:2018dy}. Considering a simple two-level system as a photon detector, the energy of the detected photon corresponds to the spacing between the energy levels. According to the standard dynamical Coulomb blockade theory \cite{Devoret_effect_of_the_electromagnetic_1990,Girvin_quantum_fluctuations_1990,Ingold_charge_1992}, the transition rates between the two levels are proportional to the so-called $P(E)$-function which characterizes the probability of the tunneling electrons to exchange energies with electromagnetic environment. Moreover, it is found that this $P(E)$-function is related to the non-symmetrized noise \cite{Schoelkopf_qubits_as_2003}. Thus, by analyzing the noise spectrum, information about light emission can also be obtained. Intuitively, one would expect that the energies of the emitted photons are limited by the applied voltage as there is a clear cutoff in the noise spectrum at a frequency equal to the applied voltage. The origin of this hard cutoff can be traced back to the Pauli exclusion principle which states that the states in the filled Fermi sea are blocked \cite{Gavish_detection_of_quantum_2000}. However, experiments in metallic tunnel contact formed by a scanning tunneling microscope (STM) show that an unexpected light emission with energies larger than the bias can be observed \cite{Schull_electron_plasmon_2009,Schneider_optical_probe_2010}. This observation was subsequently termed overbias emission and cannot be explained by a single electron process involving one photon according to the standard DCB theory \cite{Devoret_effect_of_the_electromagnetic_1990,Girvin_quantum_fluctuations_1990} which takes only the Gaussian fluctuations into consideration. Furthermore, the interaction with a local plasmon-polariton mode has been shown to be important for the overbias emission \cite{Schull_electron_plasmon_2009}. Taking the non-Gaussian fluctuations into account\cite{Tobiska_quantum_tunneling_2006}, this overbias light emission could be successfully explained by two-electron processes interacting with the local plasmon-polariton mode \cite{Xu_overbias_2014,Kaasbjerg_theory_of_light_emission_2015,Xu_dynamical_2016}. More recently even multi-electron processes have been observed experimentally in full agreement with the theory \cite{Peters_quantum_coherent_2017}.

So far most studies concerned dc-driven systems. Though in the field of time-dependent quantum transport, the ac-conductance and ac-driven noise have been widely discussed \cite{Lesovik_noise_in_an_ac_1994,Kozhevnikov_observation_of_photon_2000,Reydellet_quantum_partition_noise_2003,Rychkov_photon_assisted_2005,Safi_ac_conductance_2008,Hammer_quantum_noise_2011} and studied in the field of electron quantum optics \cite{Anonymous:3oHBDfxR,Dubois:2013dv}. Light emission from tunnel junctions at low temperatures has been experimentally addressed \cite{Gabelli:2013vr,Thibault:2014ux,Forgues:2014tf,Gasse:2013jm}. There is, however, still a lack of the general formula for the quantum noise of a contact driven by an arbitrarily shaped voltage, even without considering the light emission. Actually, ac-voltage drives will greatly affect the transport properties of the system and, thus, is expected to modify the  light emission. Hence, it is of fundamental interest to investigate the light emission in ac-driven systems and in the present article we develop an appropriate theory for the quantum noise, the DCB theory and light emission for ac-driven systems.

The rest of the article is organized as follows. In Sec.~2, we derive the general formula for the asymmetric quantum current noise for systems driven by an arbitrarily shaped periodic voltage. We discuss the influence of the shape of the drive on the quantum noise spectral density. In Sec.~3, we formulate the theory of light emission in a tunnel junction using the Keldysh path integral formalism. This theory comprises the interaction with a local plasmon-polariton mode. The Gaussian and non-Gaussian light emission rates are discussed in detail. A summary is given in Sec.~4.
%
%
\section{Non-symmetrized quantum noise for an ac-driven tunnel junction}\label{sec2}
Using the scattering matrix approach, the current operator is written as ($\hbar$=1) \cite{Buttiker_scattering_theory_1992,Pedersen_scattering_theory_1998}
%
\begin{eqnarray}
\hat{I}_\alpha(t)=\frac{e}{2\pi}\int dE dE'e^{i(E-E')t}[\hat{\bm{a}}^\dag_\alpha(E)\hat{\bm{a}}_\alpha(E')-\hat{\bm{b}}^\dag_\alpha(E)\hat{\bm{b}}_\alpha(E')],
\end{eqnarray}
%
where $\hat{\bm{a}}_\alpha$ and $\hat{\bm{b}}_\alpha$ are vectors of annihilation operators of incoming and outgoing waves in lead $\alpha$. The incoming and outgoing waves are related by the scattering matrices $\hat{\bm{b}}_\alpha(E)=\sum_\beta \bm{s}_{\alpha\beta}(E)\hat{\bm{a}}_\beta(E)$.
We assume a reservoir $\alpha$ is biased by an arbitrarily shaped time-dependent drive $\tilde{V}_\alpha(t)=\bar{V}_\alpha+\Delta V_\alpha(t)$ with period $ \omega_{ac} $. The dc-components is denoted by $\bar{V}_\alpha$ and defined by $ \bar{V}_\alpha=(\omega_{ac}/2\pi)\int_0^{2\pi/\omega_{ac}} \tilde{V}_\alpha(t)dt $. The ac-component is $\Delta V_\alpha(t)$ and obeys $ \int_0^{2\pi/\omega_{ac}} \Delta V_\alpha(t) dt=0 $. The wavefunction for the single-particle Schr\"{o}dinger equation can then be written as
%
\begin{eqnarray}
\Psi_{\alpha,n}(t,E)=\phi_{\alpha,n}(E)e^{-i(E+e\bar{V}_\alpha)t}\sum_k \cg_k\Big(\frac{eV_\alpha}{\omega_{ac}}\Big)e^{-ik\omega_{ac}t}\,.
\end{eqnarray}
%
Here $\phi_{\alpha,n}$ is the wavefunction in the transverse channel $n$ in contact $\alpha$ without the effect of the drive $\tilde{V}_\alpha(t)$ and $e^{-i\varphi_{ac}^\alpha(t)}=\sum_k \cg_k (eV_\alpha/\omega_{ac}) e^{-ik\omega_{ac}t}$ and the ac phase $\varphi_{ac}^\alpha(t)=\int_0^t dt' e\Delta V_\alpha(t')$ with $V_\alpha$ ($\omega_{ac}$) being the amplitude (frequency) of the ac component $\Delta V_\alpha(t)$ respectively. 
Similarly, the annihilation operator $\hat{\bm{a}}_\alpha$ for an incoming state close to the conductor can be expressed in terms of the reservoir states $\hat{\bm{a}}'_\alpha$ as
%
\begin{eqnarray}\label{eq:a_alpha}
\hat{\bm{a}}_\alpha(E)=\sum_k \cg_k(eV_\alpha/\omega_{ac})\hat{\bm{a}}'_\alpha(E-k\omega_{ac}).
\end{eqnarray}
%
Now the current operator becomes
%
\begin{eqnarray}
\hat{I}_\alpha(t)&=\frac{e}{2\pi}\int dE\int dE'e^{i(E-E')t}\sum_{\gamma\delta}\bm{A}_{\gamma\delta}(\alpha,E,E') \nonumber \\
&\times\sum_{kk'}\cg^*_k(\frac{eV_\gamma}{\omega_{ac}})\cg_{k'}(\frac{eV_\delta}{\omega_{ac}})(\hat{\bm{a}}_{\gamma}')^\dagger(E-k\omega_{ac})\hat{\bm{a}}_\delta'(E'-k'\omega_{ac}),
\end{eqnarray}
%
where we have introduced the current matrix $\bm{A}_{\gamma\delta}(\alpha,E,E')=\delta_{\alpha\gamma}\delta_{\alpha\delta}\boldsymbol{1}_\alpha-\bm{s}^\dag_{\alpha\gamma}(E)\bm{s}_{\alpha\delta}(E')$.
The average current $\langle\hat{I}_\alpha(t)\rangle$ can be calculated by taking the average of the operators $\langle(\hat{\bm{a}}_\alpha')^\dagger(E)\hat{\bm{a}}_\beta'(E')\rangle=\delta_{\alpha\beta}\delta(E-E')f_\alpha(E)$ 
with $f_\alpha(E)=[\exp((E-\mu_{\alpha})/T_e)+1]^{-1}$ being the Fermi function with chemical potential $\mu_{\alpha}$. 
The non-symmetrized noise spectrum can be calculated using the relation between the noise spectrum and the current-current correlation function $2\pi S_{\alpha\beta}(\Omega)\delta(\Omega+\Omega')=\langle\Delta\hat{I}_\alpha(\Omega)\Delta\hat{I}_\beta(\Omega')\rangle$, where $\Delta\hat{I}_\alpha(t)=\hat{I}_\alpha(t)-\langle\hat{I}_\alpha(t)\rangle$ and $\Delta\hat{I}_\alpha(\Omega)$ is the Fourier transform of $\Delta\hat{I}_\alpha(t)$, and using the Wick's theorem. Finally, the non-symmetrized noise spectra can be written as
%
\begin{eqnarray}
&S_{\alpha\beta}(\Omega)=\frac{e^2}{2\pi}\sum_{\gamma_1\gamma_2 k_1 k_2 k_3} \cg^*_{k_1}(\frac{eV_{\gamma_1}}{\omega_{ac}}) \cg_{k_2}(\frac{eV_{\gamma_2}}{\omega_{ac}})\cg_{k_3}(\frac{eV_{\gamma_1}}{\omega_{ac}}) \nonumber \\
& \times\cg^*_{k_2+k_3-k_1}(\frac{eV_{\gamma_2}}{\omega_{ac}})\int dE \ \mathrm{Tr}\Big[\bm{A}_{\gamma_1\gamma_2}(\alpha,E,E-\Omega) \nonumber \\
& \times \bm{A}_{\gamma_2\gamma_1}(\beta,E-\Omega+(k_3-k_1)\omega_{ac},E+(k_3-k_1)\omega_{ac})\Big] \nonumber \\
& \times f_{\gamma_1}(E-k_1\omega_{ac})(1-f_{\gamma_2}(E-\Omega-k_2\omega_{ac})).
\end{eqnarray}
%
\\
Consider a 2-terminal conductor with the scattering matrices independent of energy. The autocorrelation noise can be written as ($k_B=1$)
%
\begin{eqnarray}\label{eq:SLL}
S_{LL}(\Omega)=\frac{e^2T_e}{2\pi}&& \Bigg\{2\sum_n T^2_n\Theta^{ns}(\frac{\Omega}{2T_e}) 
+\sum_n T_n(1-T_n) \sum_k |\cg_k(N)|^2 \nonumber \\
&&\times\Big[\Theta^{ns}(\frac{\Omega+e\bar{V}+k\omega_{ac}}{2T_e}) 
+\Theta^{ns}(\frac{\Omega-e\bar{V}-k\omega_{ac}}{2T_e})\Big] \Bigg\},
\end{eqnarray}
%
where $\Theta^{ns}(x)=xe^{-x}\csch(x)$, $N=eV_0/\omega_{ac}$, 
$V_0=V_L-V_R$, 
$\bar{V}=\bar{V}_L-\bar{V}_R$, $T_e$ is the temperature and $\{T_n\}$ are the energy-independent transmission eigenvalues. Above we have used the integration
%
\begin{eqnarray}
\int^\infty_0 dE f_2(E)(1-f_1(E))=\frac{\mu_1-\mu_2}{2\sinh(\frac{\mu_1-\mu_2}{2T_e})}\exp\Big(-\frac{\mu_1-\mu_2}{2T_e}\Big) \, . 
\end{eqnarray}
%
The above formula can also be generalized to symmetrized noise. For symmetrized noise $\Theta^s(x)=x\coth(x)$ and its relation to non-symmetrized case is $\Theta^{ns}(x)=\Theta^s(x)-x$. 
In this paper, we are mainly interested in the non-symmetrized noise of the tunnel junction $(T_n \ll 1)$. 
In the following $S_{LL}$ will be the non-symmetrized noise for the tunnel junction without further clarification and it is simplified 
as $S_{LL}=\left[ (e^2\omega_{ac}/2\pi)\sum_nT_n \right] \tilde{S}_{LL}$. Here we have introduced the dimensionless non-symmetrized noise
%
\begin{eqnarray}
\tilde{S}_{LL}(\Omega)=\frac{\cW_{ac}(\Omega)-\Omega}{\omega_{ac}}, \ \ \cW_{ac}(x)=\frac{1}{2}\left[W_{ac}(x)+W_{ac}(-x)\right]
\end{eqnarray}
%
where $W_{ac}(x)=\sum_k |\cg_k|^2 W(x-k\omega_{ac}-e\bar{V})$ with  $W(x)=x\coth(x/2T_e)=2T_e\Theta^s(x/2T_e)$. At low temperature $T_e\ll\omega_{ac}$, this $W$-function reduces to $W(x)\approx|x|$. In this paper, we mainly consider the low temperature case where shot noise dominates.
%
\begin{figure}[htp]
	\centering
	\includegraphics[scale=0.9]{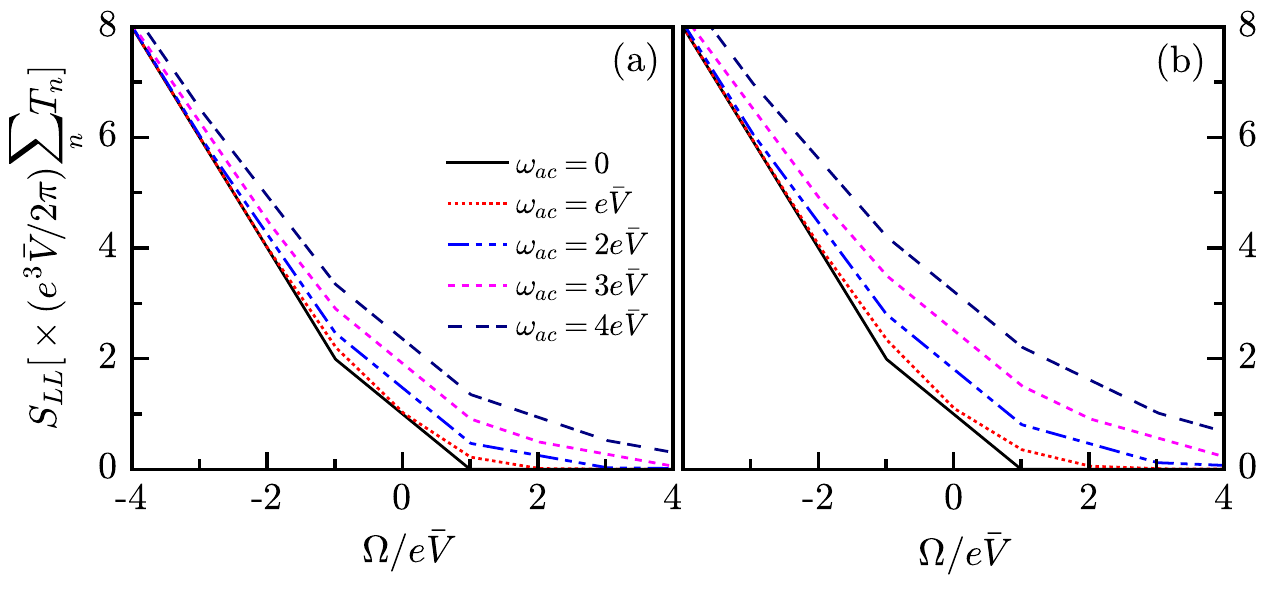}
	\caption{\label{fig:nsym-noise}Non-symmetrized noise of the tunnel junction for different applied voltages $V(t)=\bar{V}+\Delta V(t)$ with period $\tau=2\pi/\omega_{ac}$ at low temperature. (a) For harmonic drive with $\Delta V(t)=V_0\cos(\omega_{ac}t)$. (b) For square-wave drive with $\Delta V(t)=V_0$ for $0<t<\tau/2$ and $\Delta V(t)=-V_0$ for $\tau/2<t<\tau$. Here $N=eV_0/\omega_{ac}=1$. }
\end{figure}
%
In figure~\ref{fig:nsym-noise} the non-symmetrized noises with different ac frequencies $\omega_{ac}$ for harmonic and square-wave drives \cite{Vanevic_elementary_events_2007,Vanevic_elementary_charge_2008} are shown. Note that the noise spectrum with pure dc drive ($V_0=0$) exhibits a clear cutoff at $\Omega=e\bar{V}$. The inclusion of ac drive smears out this feature.
%
\begin{figure}[htp]
	\centering
 	\includegraphics[scale=0.65]{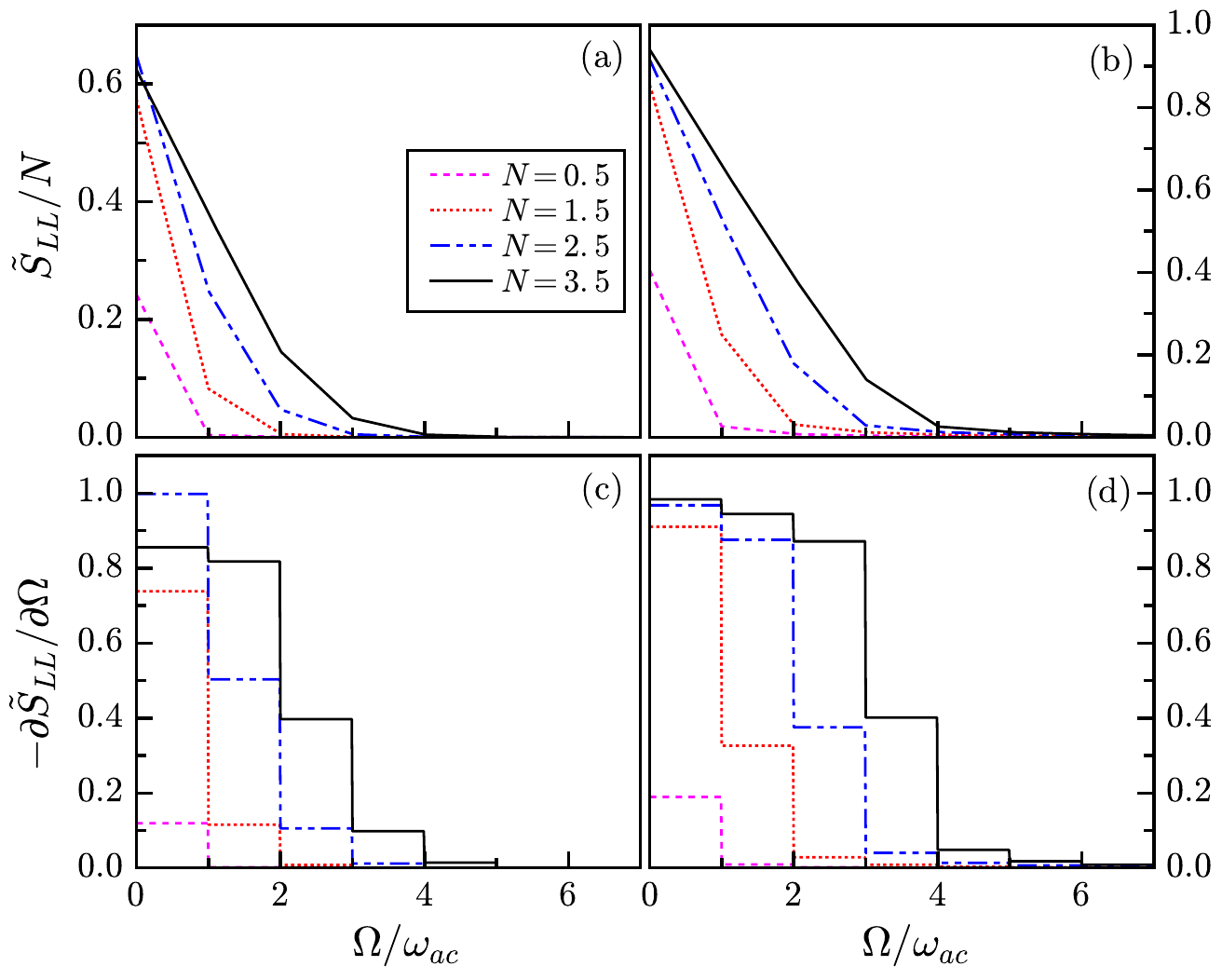}
	\caption{\label{fig:diff-noise}(a) (b) Scaled dimensionless non-symmetrized noises $\tilde{S}_{LL}/N$ and (c) (d) their corresponding differential noises $\partial\tilde{S}_{LL}/\partial\Omega$ for (a) (c) harmonic drive and (b) (d) square-wave drive. The dc voltage is zero $\bar{V}=0$.}
\end{figure}
%
In figure~\ref{fig:diff-noise}, the dimensionless non-symmetrized noises $\tilde{S}_{LL}$ and their corresponding differential noises $\partial\tilde{S}_{LL}/\partial\Omega$ for pure ac drives ($\bar{V}=0$) are shown. The noise spectra for both harmonic and square-wave drives display kinks at $\Omega=n\omega_{ac}$ ($n=1,2,3,\cdots$), which can be explained by the photon-assisted tunneling \cite{Tien_multiphoton_process_1963}. This becomes more clear if we look at the differential noises. The differential noises are piecewise constants as a function of $\Omega$. At $\Omega/\omega_{ac}=n$, $n$ ac-excited photons are created and contribute to the electron transport. With the increase of the ratio between the ac amplitude $V_0$ and ac frequency $\omega_{ac}$, namely $N$, the number of the ac-excited photon is also increased, which in the end leads to the increase of the noises. In figure~\ref{fig:diff-noise} we consider the case with pure ac drive, while it is straightforward to see from equation~(\ref{eq:SLL}), with finite dc voltage applied, the kinks of the noise spectra will be shifted to $\Omega=n\omega_{ac}+e\bar{V}$.
%
\begin{figure}[htp]
	\centering
	\includegraphics[scale=0.65]{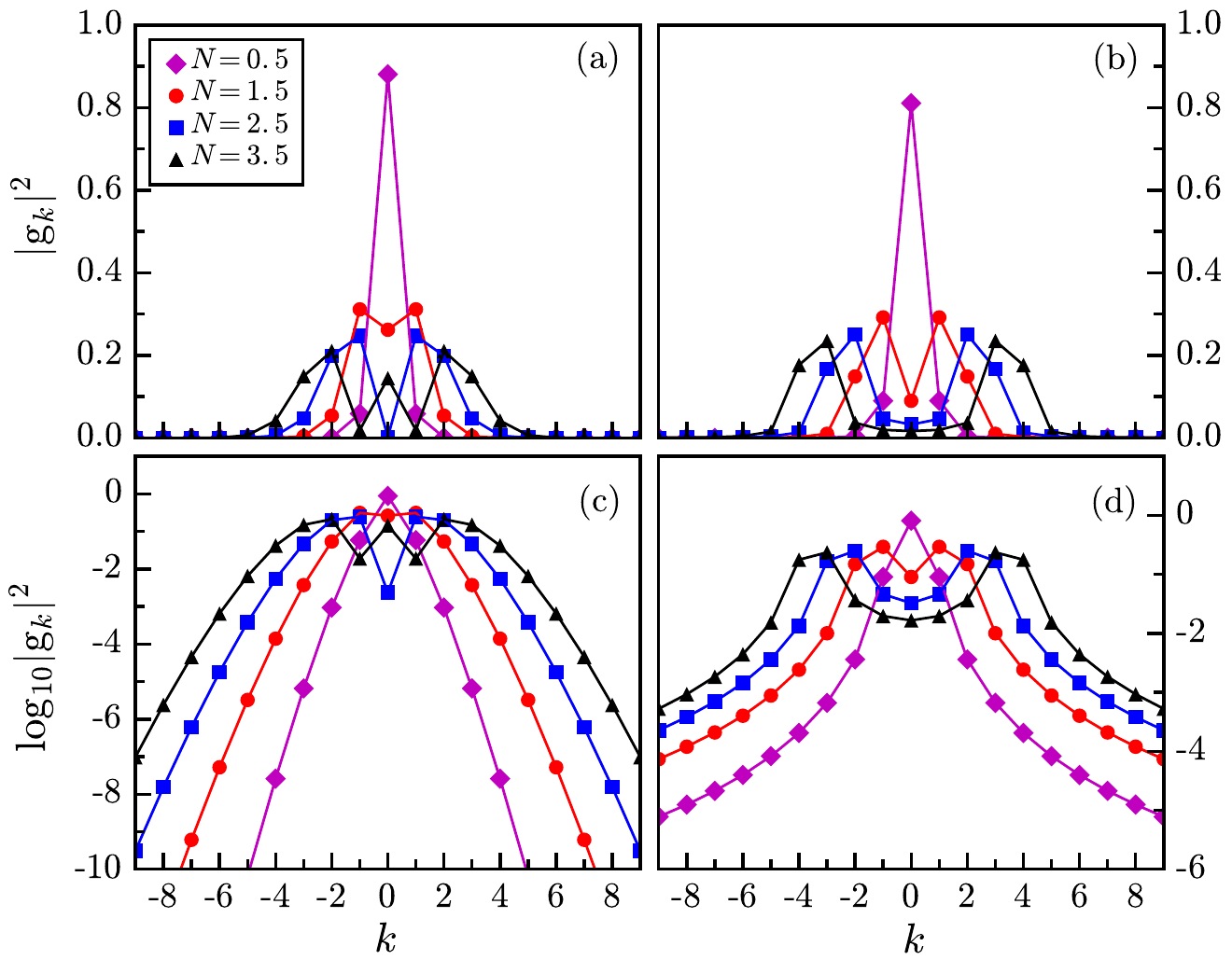}
	\caption{\label{fig:gk-distribution}Distribution of $|\cg_k|^2$ and $\log_{10}|\cg_{k}|^2$ for (a) (c) harmonic drive and (b) (d) square-wave drive for different $N$.}
\end{figure}
%
%
\begin{figure}[htp]
	\centering
	\includegraphics[scale=0.9]{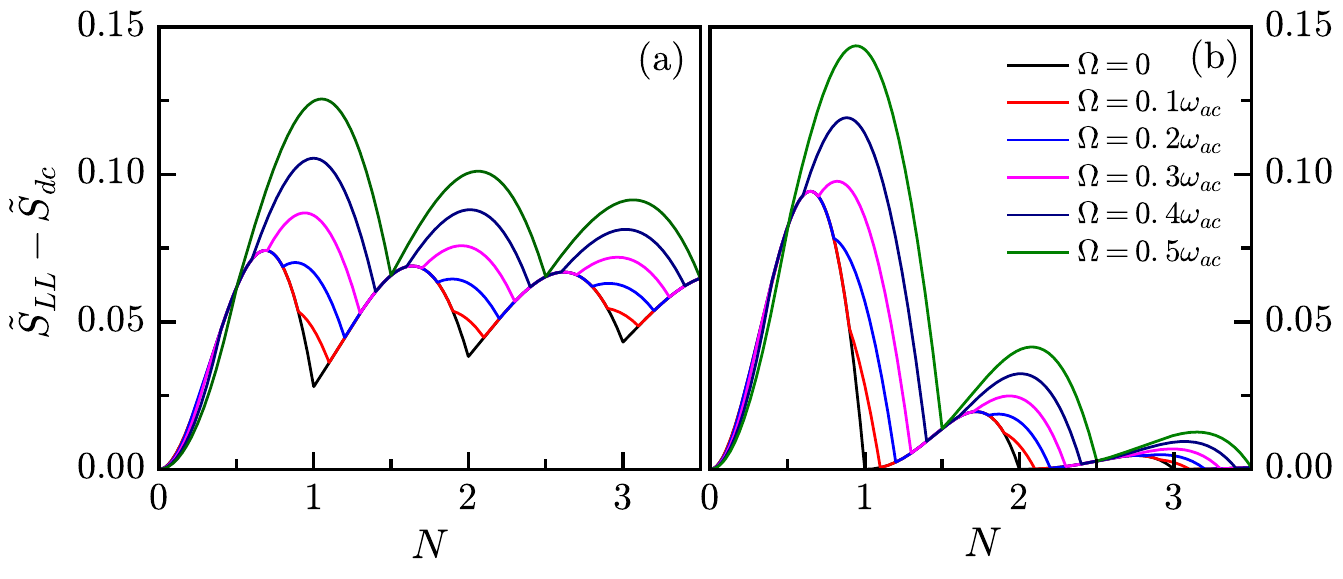}
	\caption{\label{fig:excess-noise}Dimensionless excess noises $\tilde{S}_{LL}-\tilde{S}_{dc}$ for (a) harmonic drive and (b) Lorentzian drive of width $\tau_L$ with $\Delta V(t)=-V_0+\frac{V_0}{\pi}\sum_{k=-\infty}^{\infty}\frac{\tau\tau_L}{(t-k\tau)^2+\tau_L^2}$ at low temperature. The amplitude and the dc voltage increase simultaneously, $V_0=\bar{V}$.}
\end{figure}
%
Besides, the noise spectra are also affected by the shape of the ac drive through the Fourier coefficients $\cg_k$ which satisfy
%
\begin{eqnarray}\label{eq:sumgk}
\sum_{k=-\infty}^{\infty}\cg_{n+k}\cg_{m+k}^*=\delta_{nm}, \ \ \ \sum_{k=-\infty}^{\infty}k|\cg_{k}|^2=0.
\end{eqnarray}
%
Thus $|\cg_{k}|^2$ can be regarded as the probabilities for electron to absorb ($k>0$) or emit ($k<0$) $|k|$ ac-excited photons. For better illustration, $|\cg_k(N)|^2$ and their logarithms for different $k$ and $N$ are plotted in figure~\ref{fig:gk-distribution}. For $N\ll1$, $|\cg_{0}|^2$ dominates. With the increase of $N$, higher harmonics of $|\cg_{k}|^2$ ($k=\pm1,\pm2,\cdots$) start to contribute. One thing to note, for harmonic drive and square-wave drive, $|\cg_{k}|^2=|\cg_{-k}|^2$ for arbitrary $k$. 
For Lorentzian drive, there is no such relation, i.e. $|\cg_{k}|^2\neq|\cg_{-k}|^2$. 
However, equation~(\ref{eq:sumgk}) is valid for any periodic drives and thus still holds for Lorentzian drive (see caption in Fig.~ \ref{fig:excess-noise}).
The differential dimensionless noises with zero dc voltage as shown in figure~\ref{fig:diff-noise} for $\Omega>0$ can be written as $\partial\tilde{S}_{LL}/\partial(-\Omega)=(1-\sum_{k=\ceil{-\Omega/\omega_{ac}}}^{\floor{\Omega/\omega_{ac}}}|\cg_k|^2)/\omega_{ac}$, where $\ceil{x}$ is the smallest integer greater than or equal to $x$ and $\floor{x}$ is the largest integer less than or equal to $x$. From this we can see there is a direct connection between the differential noises and $|\cg_{k}|^2$. For example, the differential noises for harmonic drive with $N=3.5$ change significantly at $\Omega/\omega_{ac}=2, 3$ while for square-wave drive, they significantly change at $\Omega/\omega_{ac}=3, 4$ as shown in figure~\ref{fig:diff-noise}, which corresponds to the facts that the leading orders of $|\cg_{k}|^2$ with $N=3.5$ for harmonic drive are $|\cg_{\pm2}|^2$ and $|\cg_{\pm3}|^2$ and for square-wave drive are $|\cg_{\pm3}|^2$ and $|\cg_{\pm4}|^2$ as shown in figure~\ref{fig:gk-distribution}.

For zero-frequency noise spectra, it is known that the Lorentzian drive carrying integer number of charge quanta $N=n$ ($n=1,2,3,\cdots$) exhibits zero excess noise \cite{Levitov_electron_counting_1996,Keeling_minimal_excitation_2006,Dubois_minimal_2013,Dubois_integer_2013}. Here we could also define the dimensionless finite-frequency excess noise $\tilde{S}_{LL}-\tilde{S}_{dc}$ and $\tilde{S}_{dc}(\Omega)=[W(\Omega+e\bar{V})
+W(\Omega-e\bar{V})-2\Omega]/2\omega_{ac}$. In figure~\ref{fig:excess-noise}, the finite-frequency excess noises for harmonic drive and Lorentzian drive are shown. Unlike zero-frequency excess noise, the locations of the minimum of the finite-frequency excess noises are shifted to $N=n+\Omega/\omega_{ac}$. Extra kinks appear at $N=n-\Omega/\omega_{ac}$. Furthermore, for Lorentzian drive, the finite-frequency excess noises do not vanish at minimum, which is due to the fact that the finite-frequency excitations destroy the ideal integer charge quanta situation.
%

%
\section{Dynamical Coulomb blockade theory of light emission from a tunnel junction}
%
\subsection{Model}
\begin{figure}[htp]
	\centering
	\includegraphics[scale=0.6]{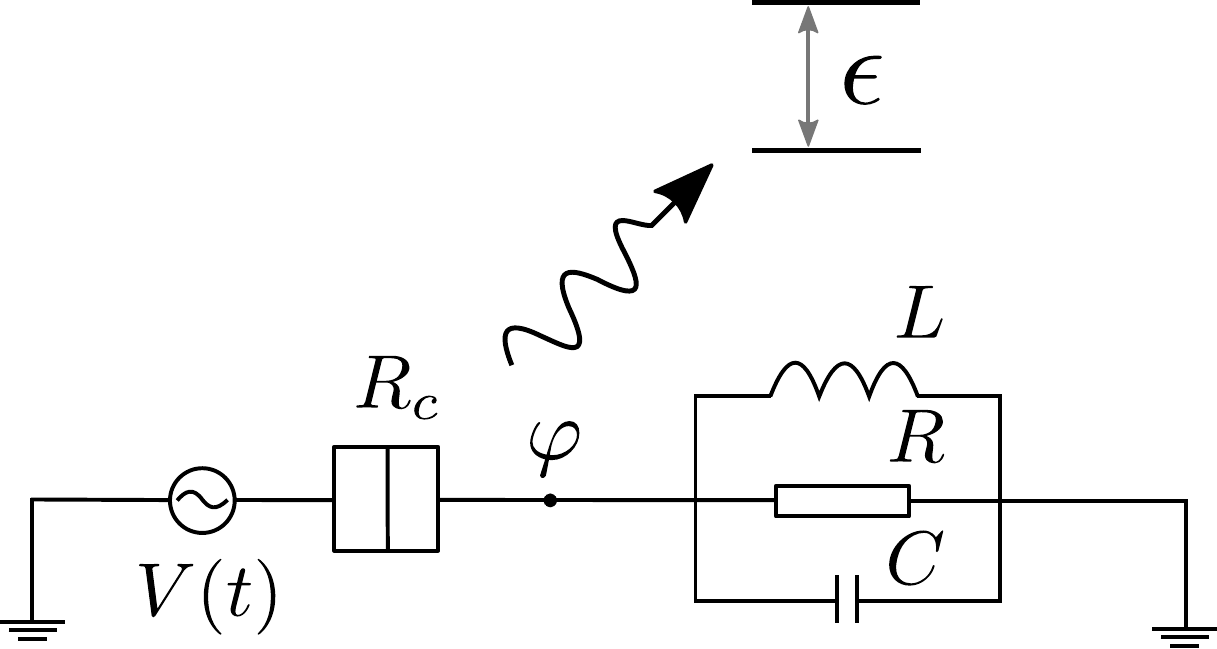}
	\caption{\label{fig:setups}Schematic setup for light emission from a STM junction coupled to a single resonant mode (SRM) (see \cite{Xu_overbias_2014,Xu_dynamical_2016}). Here the STM junction is modeled as a tunnel junction while the SRM is depicted by a LRC resonant circuit. The emitted photons are captured by the photon detector which is assumed to be a two-level system.}
\end{figure}

It is experimentally found that light emission occurs not only at energy lower than the dc bias but also at the overbias energy $\Omega>e\bar{V}$ \cite{Schull_electron_plasmon_2009}. This overbias emission is successfully explained by  two-electron and even multi-electron processes \cite{Xu_overbias_2014,Xu_dynamical_2016,Peters_quantum_coherent_2017} emitting a single photon. However, light emission with time-dependent drive has not been investigated so far. Since the transport property is remarkably altered by the ac drive compared with the dc drive as shown in Sec.~\ref{sec2}, it is of great interest to investigate the light emission from the ac-driven system. Here the light emission from a STM is modeled by a tunnel junction with dimensionless conductance $g_c = R_Q/R_c=\sum_nT_n$ and $R_Q =\pi/e^2$, $R_c$ being the quantum and tunneling resistances, respectively. The tunnel junction is coupled to a LRC resonant circuit with impedance $z_\omega=iz_0\omega\omega_0/(\omega_0^2-\omega^2+ i\omega\eta)$, $\omega_0 = 1/\sqrt{LC}$ being the frequency of the single resonant mode (SRM), $\eta = 1/RC$ being the damping, and $z_0 = \sqrt{L/C}/R_Q$ being the scaled characteristic impedance, as shown in figure~\ref{fig:setups}. The current fluctuations in the tunnel junction are transformed into voltage fluctuations $\delta V(t)$ on the node between the tunnel junction and the LRC circuit. Mathematically, the dynamical voltage is expressed as the fluctuating phase $\varphi(t) =e\int_{-\infty}^t dt'\delta V(t')$. 
%
%
The photon detector is modeled as a two-level system for simplicity, with the level spacing $\epsilon$ \cite{Schoelkopf_qubits_as_2003}. 
The tunnel amplitude between the two level is $\cT$ and is modified as $\cT e^{i\alpha\varphi(t)}$ in the presence of voltage fluctuations. 
Here $\alpha$ is the coupling constant between the energy levels and the voltage fluctuations and is assumed to be weak since in most experiments the photon detectors are far away from the tunnel junctions \cite{Xu_overbias_2014,Xu_dynamical_2016}.

The transition rate between the two levels due to the voltage fluctuations can be calculated using the Fermi's golden rule \cite{Ingold_charge_1992}
%
\begin{eqnarray}
\Gamma(\epsilon)=|\cT|^2\int dt\langle e^{i\alpha\varphi(t)}e^{-i\alpha\varphi(0)}\rangle e^{-i\epsilon t}.
\end{eqnarray}
%
In this work we study only the absorption rate of the detector, namely the emission rate of the tunnel junction, which corresponds to $\epsilon>0$. The correlator can be calculated using the path integral method and expressed as
%
\begin{eqnarray}
\langle e^{i\alpha\varphi(t)}e^{-i\alpha\varphi(0)}\rangle=\int&\cD[\Phi] \exp\{-i\cS_e[\Phi]-i\cS_c[\Phi] \nonumber \\
&+i\alpha[-\varphi^+(0)+\varphi^-(t)]\},
\end{eqnarray}
%
where $\Phi=((\varphi^+ +\varphi^-)/2,\varphi^+ -\varphi^-)^T$ and $\varphi^\pm$ are defined on the forward and backward Keldysh contours respectively. The environmental action describing the LRC circuit is given by \cite{Kindermann_interaction_effects_2003,Kindermann_Feedback_2004}
%
\begin{eqnarray}
\cS_e[\Phi]=\int \frac{d\omega}{2\pi} \Phi_{-\omega}^T A_\omega \Phi_{\omega}, A_\omega=-\frac{i}{4\pi}\left(\begin{array}{ccc}
0 & -\frac{\omega}{z_{-\omega}} \\
\frac{\omega}{z_\omega} & W(\omega)\mathrm{Re}\{\frac{1}{z_\omega}\}
\end{array}\right).
\end{eqnarray}
%
\\
The conductor action describing the tunnel junction $\cS_c$ can be expressed \cite{Nazarov_novel_circuit_1999,Snyman_keldysh_action_2008} in terms of Keldysh Green's function $\cG_{L,R}$
%
\begin{eqnarray}
\cS_c=\frac{i}{4} g_c\int dtdt' \mathrm{Tr}[\cG_L(t,t')\cG_R(t'-t)] \, ,
\end{eqnarray}
%
with the equilibrium Keldysh Green's function
%
\begin{eqnarray}
\cG_{eq}(\omega)=\left(
\begin{array}{ccc}
1-2f(\omega) & 2f(\omega) \\
2[1-f(\omega)] & 2f(\omega)-1
\end{array}\right)
\end{eqnarray}
%
\\
and Fermi function $f(\omega)=[\exp(\omega/T_e)+1]^{-1}$. The dc voltage is applied to $\cG_R$ so that $\cG_R(\omega)=\cG_{eq}(\omega-e\bar{V})$. The real fields and ac drive are applied to $\cG_L$ as $\cG_L(t,t')=\cU^\dag(t)\cG_{eq}(t-t')\cU(t')$ with the transformation matrix
%
\begin{eqnarray}
\cU(t)=e^{-i\varphi_{ac}(t)}\left(\begin{array}{ccc}
e^{-i\varphi^+(t)} & 0 \\
0 & e^{-i\varphi^-(t)}
\end{array}\right).
\end{eqnarray}
%
Because of the non-quadratic form of the conductor action $\cS_c$, 
the transition rate cannot be found exactly. 
However, by assuming weak coupling between the tunnel junction and the detector $\alpha\ll 1$ 
and small environmental impedance $g_c  {\left| z_\omega \right|}^2\ll 1$, the transition rate can be decomposed as
%
\begin{eqnarray}
\Gamma(\epsilon)=\Gamma_G(\epsilon)+\Gamma_{nG}(\epsilon)+\mathcal{O}(\alpha^2,g_c^2 {\left| z_\omega \right|}^4),
\end{eqnarray}
%
where the Gaussian rate scales as $\Gamma_G(\epsilon)\sim\Gamma_0=\pi\alpha^2|\cT|^2 g_cz_0^2/\omega_0$ while the non-Gaussian rate scales as $\Gamma_{nG}\sim\lambda\Gamma_0$ with $\lambda$ defined as $\lambda=g_cz_0^2$. In the absence of ac drive, the Gaussian and non-Gaussian rates reduce to the previous results obtained by Xu et al \cite{Xu_overbias_2014,Xu_dynamical_2016}.
%
\subsection{Gaussian rate}
\begin{figure}[htp]
	\centering
	\includegraphics[scale=0.8]{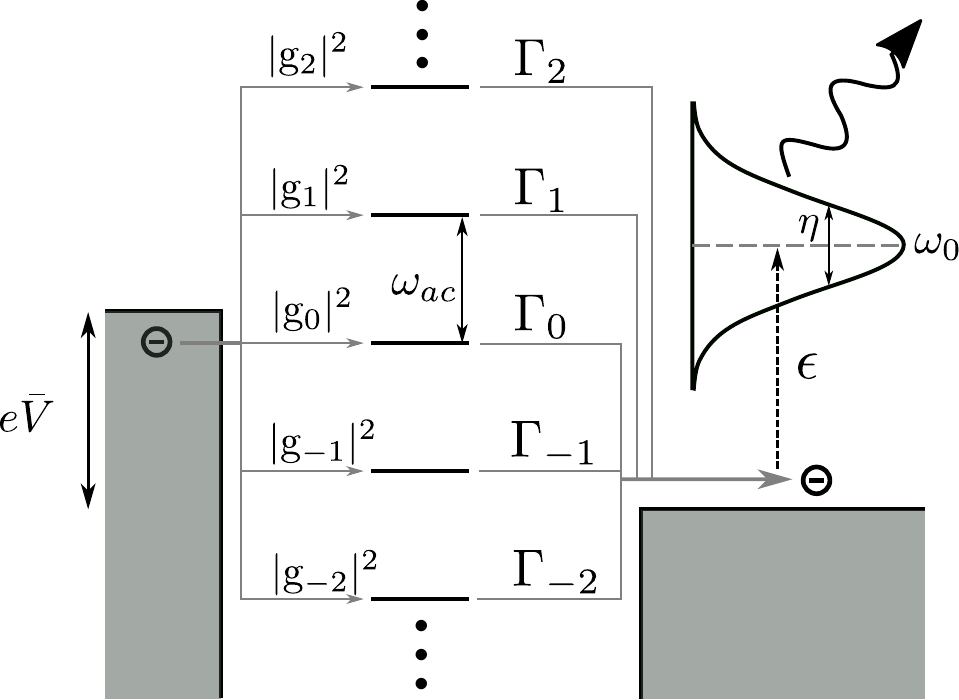}
	\caption{\label{fig:PAT-1e}Schematic diagram for one-electron tunneling process coupled to a SRM. An ac drive creates multiple sidebands and $k$ is the sideband index. Here $\Gamma_k$ represents the $k$-th electron tunneling process, where the electron can absorb/emit $k$ energy quanta (ac-excited photons) to tunnel through the barrier. $|\cg_{k}|^2$ denotes the probability of the $k$-th process.}
\end{figure}

First we only consider the Gaussian approximation, in which the action of the tunnel junction is estimated as quadratic function of the real fields. Within this approximation, $\cS_c$ is expanded up to the second order of the real fields $\varphi^\pm$. The action reads
%
\begin{eqnarray}
\cS_c^G[\Phi]&=\sum_l\int \frac{d\omega}{2\pi} \Phi_{-\omega+l\omega_{ac}}^T B_\omega^l \Phi_{\omega+l\omega_{ac}}, \ \  B_\omega^l&=-\frac{ig_c}{4\pi}\left(\begin{array}{ccc}
0 & -\omega\delta_{l0} \\
\omega \delta_{l0} & S_c^l(\omega)
\end{array}\right)
\end{eqnarray}
%
where
%
\begin{eqnarray}
S_c^l(\omega)=\frac{1}{2}\sum_k \cg_{k+l}^*\cg_{k-l}\Big[W(\omega-k\omega_{ac}-e\bar{V})
+W(\omega+k\omega_{ac}+e\bar{V})\Big]
\end{eqnarray}
%
with $S_c^0(\omega)/\omega_{ac}$ being the dimensionless symmetrized noise (see also \cite{Xu_overbias_2014,Xu_dynamical_2016}).
Within Gaussian approximation, the correlation function can be calculated by combining the environmental action and conductor action together while one should notice that the $l\neq 0$ terms vanish due to normalization condition of the path integral
%
\begin{eqnarray}\label{eq:correfunJt}
\langle e^{i\alpha\varphi(t)} e^{-i\alpha\varphi(0)} \rangle = e^{\alpha^2 J(t)} =\exp\Big\{\int\frac{d\omega}{2\pi}\frac{i}{4}\alpha^2 b^T_\omega(t) D^{-1}_\omega b_{-\omega}(t)\Big\},
\end{eqnarray}
%
where
%
\begin{eqnarray}
D_\omega=A_\omega+B_\omega^0=-\frac{i}{4\pi}\left(\begin{array}{ccc}
0 & -\frac{\omega}{\cz_{-\omega}} \\
\frac{\omega}{\cz_{\omega}} & S(\omega)
\end{array}
\right)
\end{eqnarray}
%
and $b_\omega(t)=\left(e^{-i\omega t}-1, -(e^{-i\omega t}+1)/2\right)$ with $S(\omega)=g_c S_c^0(\omega)+W(\omega)\mathrm{Re}\{1/z_\omega\}$ and $\cz_{\omega}=z_\omega/(1+z_\omega g_c)$. 
The Gaussian part of the transition rate is obtained by expanding the correlation function equation~(\ref{eq:correfunJt}) up to $\alpha^2$ and taking the Fourier transform. It is now written as
%
\begin{eqnarray}\label{eq:GammaG}
\Gamma_G(\epsilon)=2\pi\alpha^2|\cT|^2\frac{|\cz_\epsilon|^2}{\epsilon^2}S_{tot}(\epsilon),
\end{eqnarray}
%
where $S_{tot}(\epsilon)=S(\epsilon)-\epsilon\mathrm{Re}\{1/\cz_{\epsilon}\}$. For $\epsilon>0$, $S_{tot}(\epsilon)=g_c\omega_{ac}\tilde{S}_{LL}(\epsilon)$ at low temperature. 
Note that the factor $g_c$ in $\cz_{\omega}$ only leads to an increased damping of the resonator and thus can be absorbed in the renormalized damping as $\eta\rightarrow\eta+1/R_c C$. 
In the following calculation we neglect this factor because $g_c  \left| z_\omega \right| \ll 1$ and assume $\cz_{\omega}\approx z_\omega$. Equation~(\ref{eq:GammaG}) represents the light emission from the one-electron tunneling process, which can be decomposed as ($\epsilon>0$) $\Gamma_G=\sum_k\Gamma_k$ at low temperature where $\Gamma_k(\epsilon)=\pi\alpha^2|\cT|^2g_c(|\cz_\epsilon|^2/\epsilon^2)|\cg_k|^2\Big[|\epsilon-k\omega_{ac}-e\bar{V}|+|\epsilon+k\omega_{ac}+e\bar{V}|-2\epsilon\Big].$ It can be interpreted as following: An ac drive creates multiple sidebands, which leads to the tunneling of electron through the barrier by absorbing ($k>0$) or emitting ($k<0$) $k$ energy quanta with probability $|\cg_k|^2$. All these processes sum up and contribute to the excitation of the SRM and finally lead to the light emission as shown in figure~\ref{fig:PAT-1e}.
%
%
%
\begin{figure}[htp]
	\centering
	\includegraphics[scale=0.9]{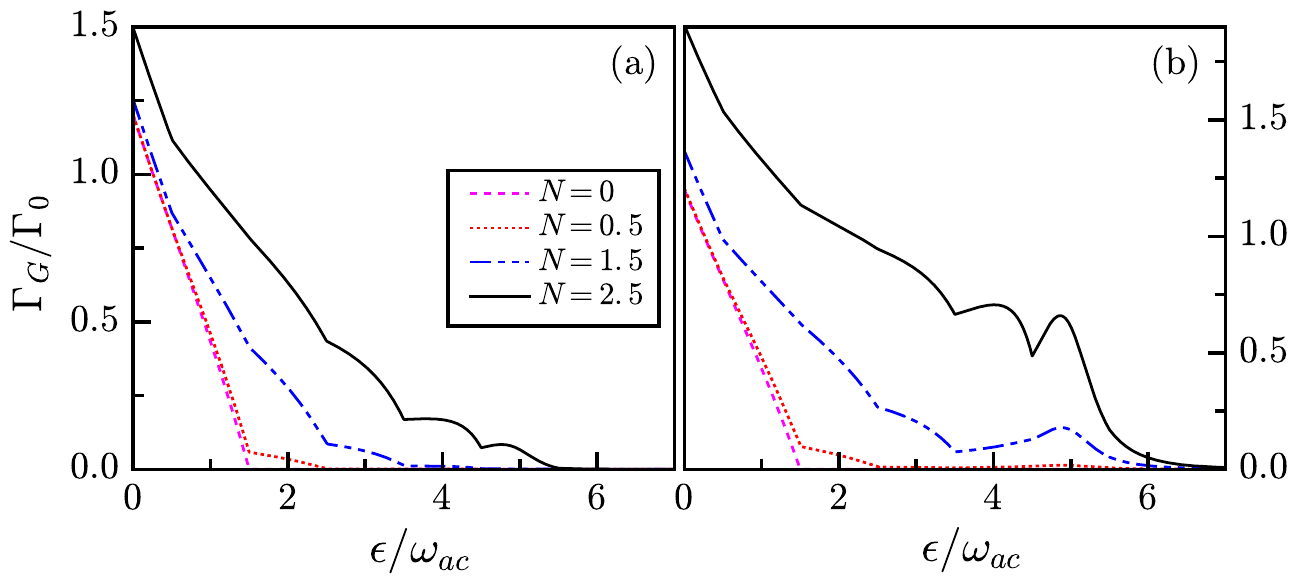}
	\caption{\label{fig:LE-eV-G}Gaussian part of the transition rates with finite dc voltage at low temperature for different $N$, for (a) harmonic drive and (b) square-wave drive. Here $e\bar{V}=1.5\omega_{ac}$, $\omega_0=5\omega_{ac}$ and $\eta=\omega_{ac}$.}
\end{figure}
%
%
\begin{figure}[htp]
	\centering
	\includegraphics[scale=1]{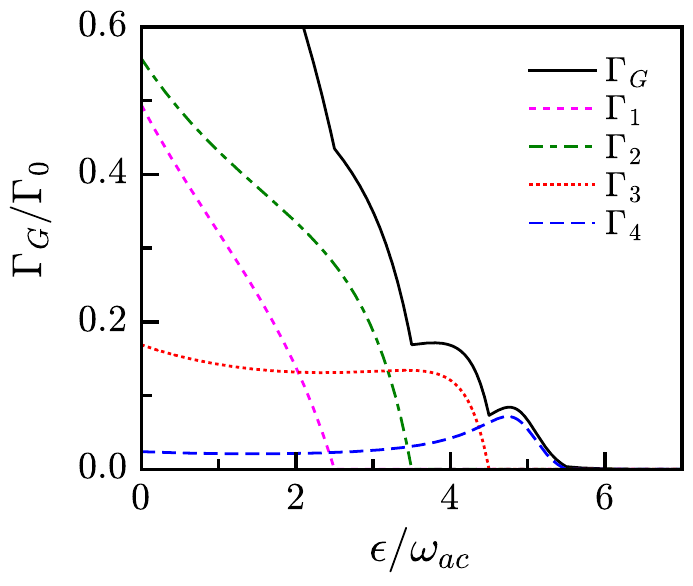}
	\caption{\label{fig:LE-eV-G-dcp}Decompositions of Gaussian part of the transition rates for harmonic drive at low temperature. Here $N=2.5$, $e\bar{V}=1.5\omega_{ac}$, $\omega_0=5\omega_{ac}$ and $\eta=\omega_{ac}$.}
\end{figure}
%
%
\begin{figure}[htp]
	\centering
	\includegraphics[scale=1]{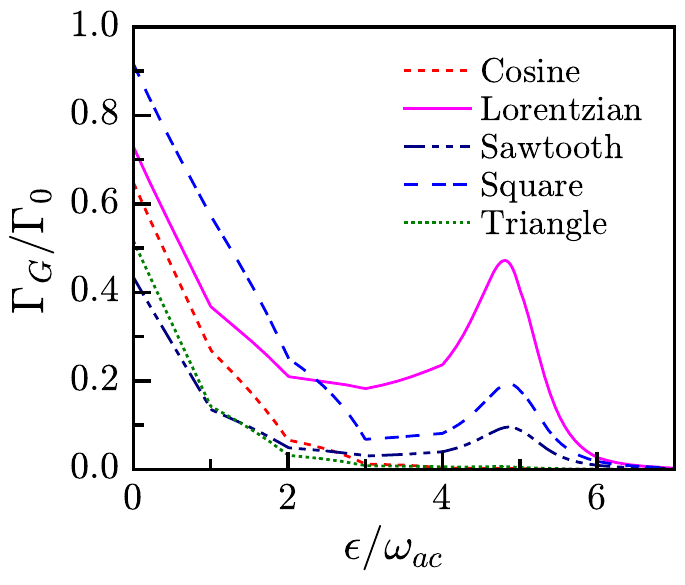}
	\caption{\label{fig:LE-N-G}Gaussian parts of the transition rates with zero dc voltage $\bar{V}=0$ at low temperature for different drives with sawtooth drive $\Delta V(t)=2V_0t/\tau-V_0$ and triangle-wave drive $\Delta V(t)=4V_0t/\tau-V_0$ for $0<t<\tau/2$ and $\Delta V(t)=-4V_0t/\tau+3V_0$ for $\tau/2<t<\tau$. Here $N=2.5$, the frequency of the SRM $\omega_0=5\omega_{ac}$ and the broadening $\eta=\omega_{ac}$.}
\end{figure}
%
%
\begin{figure}[htp]
	\centering
	\includegraphics[scale=1]{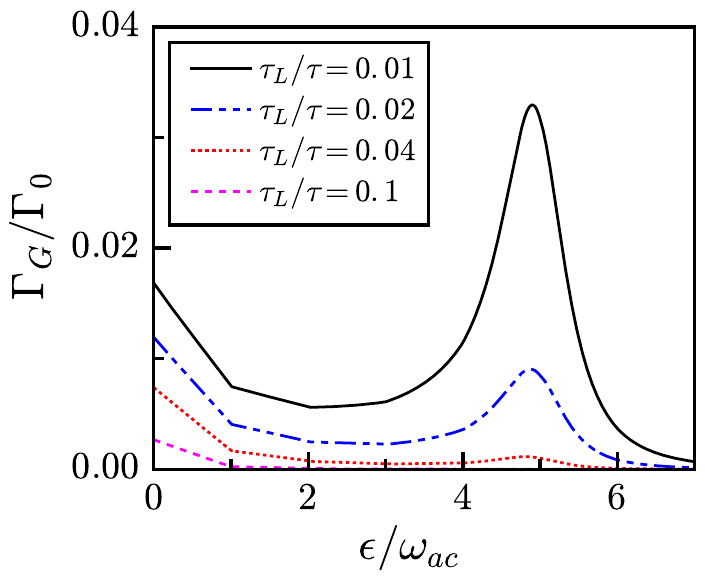}
	\caption{\label{fig:LE-N-G-Lor}Gaussian rates with zero dc voltage $\bar{V}=0$ for Lorentzian drive at low temperature for different $\tau_L/\tau$ characterizing the sharpness of the Lorentzian drive. Here $N=0.1$, the frequency of the SRM $\omega_0=5\omega_{ac}$ and the broadening $\eta=\omega_{ac}$.}
\end{figure}
%

%
In absence of the ac drive, for $e\bar{V}<\omega_0$, the Gaussian rates exhibit a clear cutoff at $\epsilon=e\bar{V}$  \cite{Xu_dynamical_2016}, 
which is  shown in figure~\ref{fig:LE-eV-G} for $N=0$.
Finite temperature effects were also analyzed in  \cite{Xu_dynamical_2016} for dc voltage bias tunnel junction.
Here we focus the analysis of the emission rate at low temperature with an arbitrary ac-drive.
Example of results for $\Gamma_G(\epsilon)$ are shown in the figures \ref{fig:LE-eV-G}, \ref{fig:LE-eV-G-dcp}, \ref{fig:LE-N-G}  and \ref{fig:LE-N-G-Lor}.
%
%
At small ac driving amplitude, e.g. $N=0.5$, the Gaussian rates exhibit small deviations from the pure dc drive case, where the cutoff at $\epsilon=e\bar{V}$ is lifted. The overbias light emission at $\epsilon>e\bar{V}$ is due to the interplay between the electrons and the ac-excited photons. As $N$ increases, the kink at $\epsilon=e\bar{V}$ is smeared out. There are extra kinks that appear at $\epsilon=e\bar{V}\pm n\omega_{ac}$. Furthermore, the Gaussian rates just mimic the features of the non-symmetrized noises at energy far away from the resonance frequency $\omega_0$, which is due to the fact that the impedance behaves like a quasilinear function at energy far away from the resonance frequency. At energy close to $\omega_0$ and with sufficiently large ac amplitude, e.g. $N=2.5$, the resonance peak of the SRM starts to appear. Here we should note that not only the driving strength but also the shape of the drive affect the behavior of the Gaussian rates. For example, for square-wave drive, there is a more pronounced peak at $\epsilon\simeq\omega_0$ than for harmonic drive for $N=2.5$ case, whose origin is the same as the discussion of the differential noises, that is the leading-order harmonics of $|\cg_k|^2$ for square-wave drive are $|\cg_{\pm2}|^2$ and $|\cg_{\pm3}|^2$ while for harmonic drive, they are $|\cg_{\pm1}|^2$ and $|\cg_{\pm2}|^2$. The physics behind is the following, for square-wave drive, the tunneling electron can absorb most likely 2 or 3 ac-excited photons, which leads to the emission of photon from the SRM with energy close to $\omega_0=5\omega_{ac}$. While for harmonic drive, the electron can only absorb most likely 1 or 2 ac-excited photons, with which the energy of the electron is not sufficiently large enough to reach the frequency of the SRM. In figure~\ref{fig:LE-eV-G-dcp}, we plot the leading-order decompositions of the emission rate for $N=2.5$, where there are clear cutoffs for each $\Gamma_k$ since $\Gamma_k\propto\theta(|k\omega_{ac}+e\bar{V}|-\epsilon)$ with $\theta(x)$ being the Heaviside step function.

In figure~\ref{fig:LE-N-G}, the Gaussian rates with zero dc voltage $\bar{V}=0$ for different shapes of drives are shown. It can be seen that the shapes of drives greatly affect the light emission. For example, with the same $V_0$ and $\omega_{ac}$, the emission rates for Lorentzian, square-wave and sawtooth drives can exhibit the resonance peaks at $\epsilon\simeq\omega_0$. While for harmonic and triangle-wave drives, this peak is hardly seen. The Lorentzian drive is special because its corresponding $|\cg_{k}|^2$ is not symmetrized over $k$ and it has a long tail compared to other drives we mentioned. 
Thus with the same ac driving strength, Lorentzian drive would be the most likely drive to observe the resonance peaks of the SRM. 
Furthermore, the resonance peak of the SRM can emerge even at small $N$ for a sharp Lorentzian drive with $\tau_L/\tau\ll 1$, as shown in figure~\ref{fig:LE-N-G-Lor}.
%
\subsection{Non-Gaussian rate}
Beyond the Gaussian approximation, the correlation function cannot be evaluated exactly. Thus we made the following approximations. First we expand the action of the conductor up to the fourth order in the real fields. The correlation function is expressed as
%
\begin{eqnarray}
\langle e^{i\alpha\varphi(t)} e^{-i\alpha\varphi(0)} \rangle \approx e^{\alpha^2 J(t)} -i\llangle\cS_c^{(3)}\rrangle -i\llangle\cS_c^{(4)}\rrangle
\end{eqnarray}
%
where the Gaussian averages of the moments are given by
%
\begin{eqnarray}\label{eq:correfun3rd4thorder}
\llangle\dots\rrangle \equiv \int\cD[\Phi](\dots)e^{\int\frac{d\omega}{2\pi}\{-i\Phi_{-\omega}^TD_\omega\Phi_{\omega} +i\alpha b_\omega^T(t)\Phi_{\omega}\}},
\end{eqnarray}
%
which can be easily evaluated by taking the derivatives of equation~(\ref{eq:correfunJt}) with respect to $b_\omega$. The results read
%
\begin{eqnarray}
\llangle\Phi_{\omega}\rrangle =\frac{\alpha}{2} D_\omega^{-1}b_{-\omega}(t), \ \ \
\llangle\Phi_{\omega}\Phi_{-\omega}^T\rrangle =-\frac{i}{2}D_\omega.
\end{eqnarray}
%
Now the non-Gaussian contribution of $\cS_c$ is written as
%
\begin{eqnarray}
\cS_c^{nG}\approx\cS_c^{(4)} 
=&\frac{ig_c}{96} \sum_{kl}\cg_{k+l}^*\cg_{k-l}\int\frac{d\omega_1}{2\pi}\frac{d\omega_2}{2\pi}\frac{d\omega_3}{2\pi} \nonumber \\
&\times\Big\{ \cS_{c1}^{nG}+\cS_{c2}^{nG}+\cS_{c3}^{nG}+\cS_{c4}^{nG}+\cS_{c5}^{nG} \Big\}
\end{eqnarray}
%
with the components expressed as
\numparts 
%
\begin{eqnarray}
\cS_{c1}^{nG}=&\varphi^+_{\omega_1+l\omega_{ac}}\varphi^+_{\omega_2+l\omega_{ac}}\varphi^+_{\omega_3+l\omega_{ac}}\varphi^+_{-l\omega_{ac}-\omega_1-\omega_2-\omega_3} \nonumber \\
&\times\big[ F_{kl}^+ -4F_1(\omega_1-k\omega_{ac})-4F_1(-\omega_1-k\omega_{ac}) \nonumber \\ &+6F_1(\omega_1+\omega_2-k\omega_{ac}+l\omega_{ac})\big],
\end{eqnarray}
%
%
\begin{eqnarray}
\cS_{c2}^{nG}=&\varphi^-_{\omega_1+l\omega_{ac}}\varphi^-_{\omega_2+l\omega_{ac}}\varphi^-_{\omega_3+l\omega_{ac}}\varphi^-_{-l\omega_{ac}-\omega_1-\omega_2-\omega_3} \nonumber \\
&\times\big[F_{kl}^- -4F_1(\omega_1-k\omega_{ac})-4F_1(-\omega_1-k\omega_{ac}) \nonumber \\ &+6F_1(\omega_1+\omega_2-k\omega_{ac}+l\omega_{ac})\big],
\end{eqnarray}
%
%
\begin{eqnarray}
\cS_{c3}^{nG}=&\varphi^-_{\omega_1+l\omega_{ac}}\varphi^+_{\omega_2+l\omega_{ac}}\varphi^+_{\omega_3+l\omega_{ac}}\varphi^+_{-l\omega_{ac}-\omega_1-\omega_2-\omega_3} \nonumber \\
&\times\left[-4F_2(-\omega_1-k\omega_{ac}) -4F_3(\omega_1-k\omega_{ac})\right],
\end{eqnarray}
%
%
\begin{eqnarray}
\cS_{c4}^{nG}=&\varphi^+_{\omega_1+l\omega_{ac}}\varphi^-_{\omega_2+l\omega_{ac}}\varphi^-_{\omega_3+l\omega_{ac}}\varphi^-_{-l\omega_{ac}-\omega_1-\omega_2-\omega_3} \nonumber \\
&\times\left[-4F_2(\omega_1-k\omega_{ac}) -4F_3(-\omega_1-k\omega_{ac})\right],
\end{eqnarray}
%
%
\begin{eqnarray}
\cS_{c5}^{nG}=&\varphi^+_{\omega_1+l\omega_{ac}}\varphi^+_{\omega_2+l\omega_{ac}}\varphi^-_{\omega_3+l\omega_{ac}}\varphi^-_{-l\omega_{ac}-\omega_1-\omega_2-\omega_3} \nonumber \\
&\times\big[6F_2(\omega_1+\omega_2-k\omega_{ac}+l\omega_{ac}) \nonumber \\ 
&+6F_3(-\omega_1-\omega_2-k\omega_{ac}-l\omega_{ac})\big]\,.
\end{eqnarray}
%
\endnumparts
Here we defined
\numparts 
%
\begin{eqnarray}
F_{kl}^\pm=&F_1(-k\omega_{ac}-l\omega_{ac})+F_1(-k\omega_{ac}+l\omega_{ac}) \nonumber \\
&+F_2(-k\omega_{ac}\pm l\omega_{ac})+F_3(-k\omega_{ac}\mp l\omega_{ac}).
\end{eqnarray}
%
\endnumparts
and  $F_i (i=1,2,3)$ as 
\numparts 
%
\begin{eqnarray}
F_1(x)=\int\frac{d\omega}{2\pi} \left[1-2f_L(\omega)\right]\left[1-2f_R(\omega+x)\right] 
=\frac{1}{\pi}W(x-e\bar{V}),
\end{eqnarray}
%
%
\begin{eqnarray}
F_2(x)=\int\frac{d\omega}{2\pi} 4f_L(\omega)\left[1-f_R(\omega+x)\right] 
=\frac{1}{\pi}\Big[(x-e\bar{V})+W(x-e\bar{V})\Big],
\end{eqnarray}
%
%
\begin{eqnarray}
F_3(x)=\int\frac{d\omega}{2\pi} 4\left[1-f_L(\omega)\right]f_R(\omega+x) 
=\frac{1}{\pi}\Big[(e\bar{V}-x)+W(x-e\bar{V})\Big].
\end{eqnarray}
%
\endnumparts
Note that the third order of the action $\cS_c^{(3)}$ is neglected here because it gives a non-vanishing result only to the order $\alpha^3$, which corresponds to the unexpected process with one-and-a-half photon  \cite{Tobiska_quantum_tunneling_2006}. 

To lowest order in $\alpha^2$, the non-Gaussian part of the transition rate is written as
%
\begin{eqnarray}\label{eq:GammaNG}
	\Gamma_{nG}(\epsilon)&=&\pi^2\alpha^2|\cT|^2g_c^2\frac{|\cz_\epsilon|^2}{\epsilon^2}\int^{+\infty}_0 \frac{d\omega}{2\pi} \nonumber \\
	&&\times\Bigg\{ \frac{|\cz_\omega|^2}{\omega^2}\left[\cW_{ac}(\omega)-W(\omega)\right]\left[-2\cW_{ac}(\epsilon)+\cW_{ac}(\omega+\epsilon)+\cW_{ac}(\omega-\epsilon)\right] \nonumber \\
	&&+2\left[\cW_{ac}(\epsilon)-W(\epsilon)\right]\frac{\mathrm{Re}\{\cz_\epsilon\}}{\epsilon}\frac{\mathrm{Re}\{\cz_\omega\}}{\omega}\left[\cW_{ac}(\omega+\epsilon)-\cW_{ac}(\omega-\epsilon)\right] \nonumber \\
	&&+2\left[\cW_{ac}(\epsilon)-W(\epsilon)\right]\frac{\mathrm{Im}\{\cz_\epsilon\}}{\epsilon}\frac{\mathrm{Im}\{\cz_\omega\}}{\omega}\Big[2W(e\bar{V})-2\cW_{ac}(\epsilon)-2\cW_{ac}(\omega) \nonumber \\ &&+\cW_{ac}(\omega+\epsilon)+\cW_{ac}(\omega-\epsilon)\Big]\Bigg\},
\end{eqnarray}
%
which takes a form analogous to equation~(12) in \cite{Xu_dynamical_2016} except that 
$\hat{W}(x)=\frac{1}{2}[W(x-e\bar{V})+W(x+e\bar{V})]$ now is modified to $\hat{W}_{ac}$.
%
\begin{figure}[htp]
	\centering
	\includegraphics[scale=0.9]{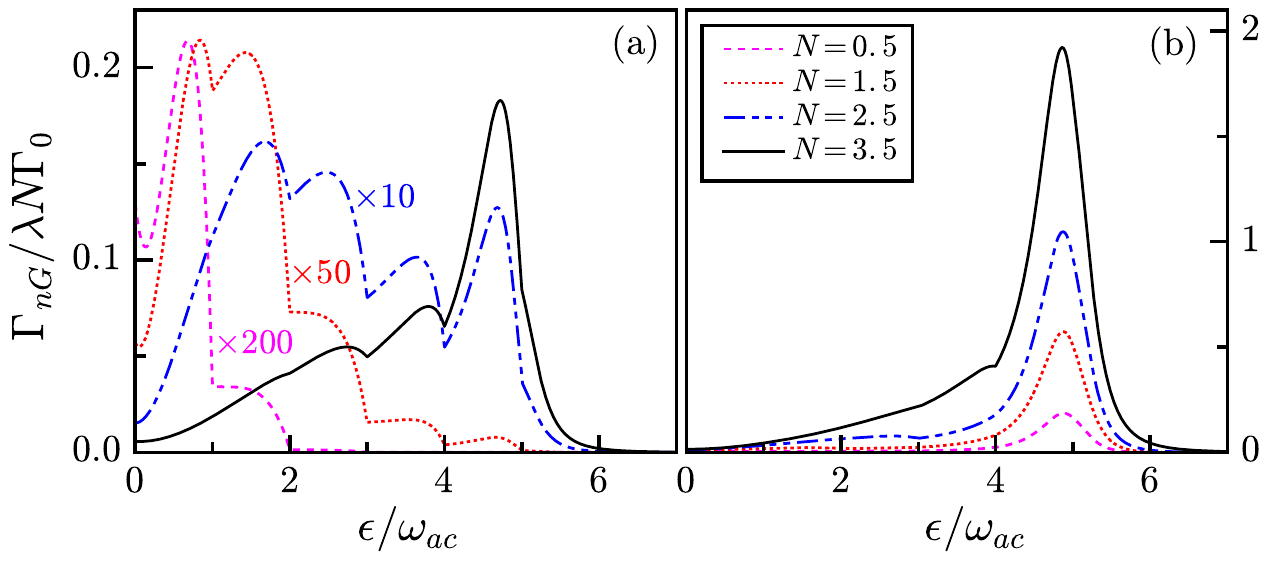}
	\caption{\label{fig:LE-N-nG}Non-Gaussian parts of the transition rates with zero dc voltage $\bar{V}=0$ at low temperature for different $N$, for (a) harmonic drive and (b) square-wave drive. Here the frequency of the SRM $\omega_0=5\omega_{ac}$ and the broadening $\eta=\omega_{ac}$.}
\end{figure}
%
While the Gaussian rates represent the one-electron tunneling process, the non-Gaussian rates describes the  two-electron tunneling process. Compared with the Gaussian rates, the non-Gaussian rates become more distinguishable between harmonic drive and square-wave drive as shown in figure~\ref{fig:LE-N-nG}. For harmonic drive, kinks at $\epsilon=n\omega_{ac}$ are clearly visible for the given $N$. As $N$ increases, the peaks are shifting towards the frequency of the SRM. With large enough $N$, the peak at $\epsilon\simeq\omega_0$ dominates and the ac-assisted tunneling peaks become invisible. For a square-wave drive, because it creates more ac-excited photons than the harmonic drive at same driving amplitude, the electrons are more likely to reach the frequency of the SRM. Thus even at small $N$, the resonance energy peak dominates. With the increase of $N$, this peak is enhanced. Noticing that with sufficiently large $N$, the satellite peaks are smeared out except the resonance energy peak. Similar behaviors appear if we increase the ac frequencies or the dc amplitudes. 
%
\begin{figure}[htp]
	\centering
	\includegraphics[scale=1]{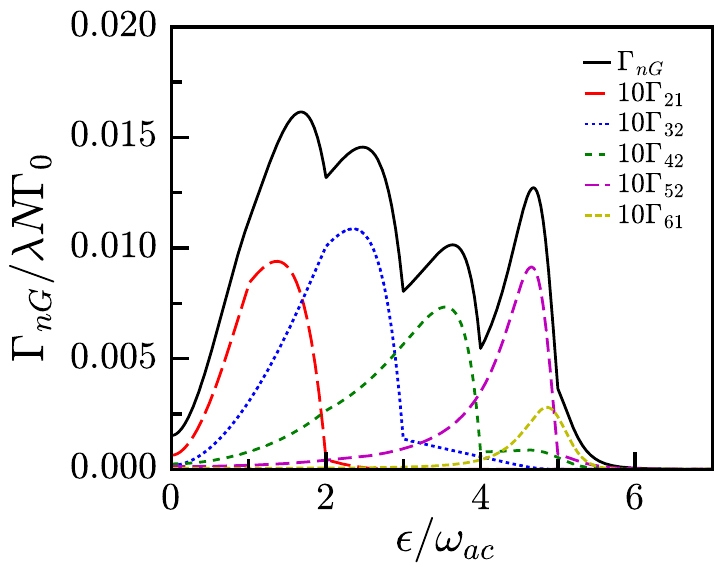}
	\caption{\label{fig:LE-N-nG-dcp}Leading-order decompositions of non-Gaussian part of the transition rates for harmonic drive at low temperature. Here $N=2.5$, $\bar{V}=0$. $\omega_0=5\omega_{ac}$ and $\eta=\omega_{ac}$.}
\end{figure}
%
%
\begin{figure}[htp]
	\centering
	\includegraphics[scale=0.7]{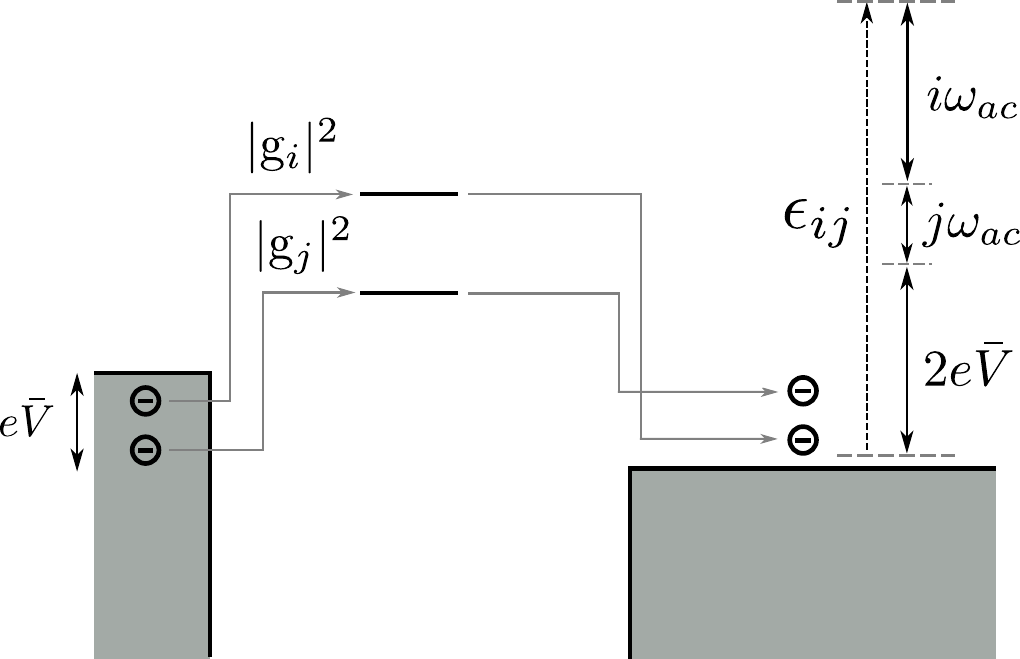}
	\caption{\label{fig:PAT-2e}Schematic diagram for two-electron tunneling process by absorbing $i(i>0)$, $j(j>0)$ energy quanta. The maximum energy of the emitted photon of this process is $\epsilon_{ij}=2e\bar{V}+(i+j)\omega_{ac}$.}
\end{figure}
%
In Figure~\ref{fig:LE-N-nG-dcp}, we plot the leading-order decompositions of the non-Gaussian rates. The decomposition reads $\Gamma_{nG}=\sum_{ij}\Gamma_{ij}$ where $\Gamma_{ij}(\epsilon)=\pi^2\alpha^2|\cT|^2g_c^2(|\cz_\epsilon|^2/\epsilon^2) |\cg_i|^2 |\cg_j|^2 \{ \Gamma_{ij}^{(1)}(\epsilon)+\Gamma_{ij}^{(2)}(\epsilon)+\Gamma_{ij}^{(3)}(\epsilon)\}$ with
\numparts 
%
\begin{eqnarray}
\Gamma_{ij}^{(1)}(\epsilon)&=&\frac{1}{4}\int^{+\infty}_0 \frac{d\omega}{2\pi} \frac{|\cz_\omega|^2}{\omega^2}\Big[W(\omega-i\omega_{ac}-e\bar{V})+W(\omega+i\omega_{ac}+e\bar{V})-2W(\omega)\Big] \nonumber \\
&&\times\Big[-2W(\epsilon-j\omega_{ac}-e\bar{V})-2W(\epsilon+j\omega_{ac}+e\bar{V}) \nonumber \\
&&+W(\omega+\epsilon-j\omega_{ac}-e\bar{V})+W(\omega+\epsilon+j\omega_{ac}+e\bar{V}) \nonumber \\ &&+W(\omega-\epsilon-j\omega_{ac}-e\bar{V})+W(\omega-\epsilon+j\omega_{ac}+e\bar{V})\Big],
\end{eqnarray}
%
%
\begin{eqnarray}
\Gamma_{ij}^{(2)}(\epsilon)&=&\frac{1}{2}\Big[W(\epsilon-i\omega_{ac}-e\bar{V})+W(\epsilon+i\omega_{ac}+e\bar{V})-2W(\epsilon)\Big]\frac{\mathrm{Re}\{\cz_\epsilon\}}{\epsilon} \nonumber \\
&&\times \int^{+\infty}_0 \frac{d\omega}{2\pi} \frac{\mathrm{Re}\{\cz_\omega\}}{\omega}\Big[ W(\omega+\epsilon-j\omega_{ac}-e\bar{V})+W(\omega+\epsilon+j\omega_{ac}+e\bar{V}) \nonumber \\
&&-W(\omega-\epsilon-j\omega_{ac}-e\bar{V})-W(\omega-\epsilon+j\omega_{ac}+e\bar{V})\Big],
\end{eqnarray}
%
%
\begin{eqnarray}
\Gamma_{ij}^{(3)}(\epsilon)&=&\frac{1}{2}\Big[W(\epsilon-i\omega_{ac}-e\bar{V})+W(\epsilon+i\omega_{ac}+e\bar{V})-2W(\epsilon)\Big]\frac{\mathrm{Im}\{\cz_\epsilon\}}{\epsilon} \nonumber \\
&&\times \int^{+\infty}_0 \frac{d\omega}{2\pi}
\frac{\mathrm{Im}\{\cz_\omega\}}{\omega}\Big[4W(e\bar{V}) -2W(\epsilon-j\omega_{ac}-e\bar{V}) \nonumber \\
&&-2W(\epsilon+j\omega_{ac}+e\bar{V})-2W(\omega-j\omega_{ac}-e\bar{V})-2W(\omega+j\omega_{ac}+e\bar{V}) \nonumber \\
&&+W(\omega+\epsilon-j\omega_{ac}-e\bar{V})+W(\omega+\epsilon+j\omega_{ac}+e\bar{V}) \nonumber \\ &&+W(\omega-\epsilon-j\omega_{ac}-e\bar{V})+W(\omega-\epsilon+j\omega_{ac}+e\bar{V}) \Big].
\end{eqnarray}
%
\endnumparts
Above we have used the condition~(\ref{eq:sumgk}). At low temperature, $W(x)\simeq |x|$ and thus $\Gamma_{ij}^{(1)}\propto\theta(|i\omega_{ac}+e\bar{V}|+|j\omega_{ac}+e\bar{V}|-\epsilon)$ while $\Gamma_{ij}^{(2),(3)} \propto\theta(|i\omega_{ac}+e\bar{V}|-\epsilon)$. Notice that in pure dc case ($V_0=0$) only $\Gamma_{ij}^{(1)}$ is responsible for the overbias light emission in \cite{Xu_overbias_2014,Xu_dynamical_2016}. The physics behind $\Gamma_{ij}$ can be explained by two-electron tunneling process as shown in figure~\ref{fig:PAT-2e}. The two electrons tunnel through the barrier by absorbing (or emitting) $i$, $j$ energy quanta with probabilities $|\cg_i|^2$, $|\cg_j|^2$, respectively. Thus the energy of the emitted photon in this process is limited by $\epsilon_{ij}=|i\omega_{ac}+e\bar{V}|+|j\omega_{ac}+e\bar{V}|$.
%
\begin{figure}[htp]
	\centering
	\includegraphics[scale=1]{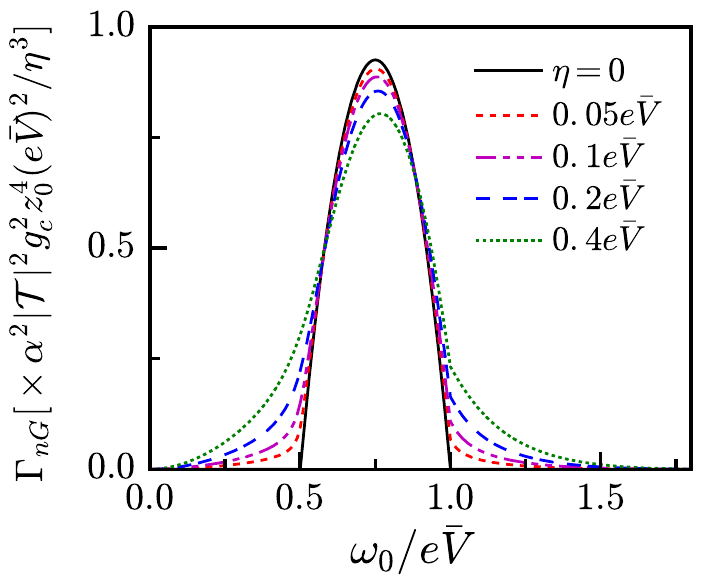}
	\caption{\label{fig:LE-nG-zero-eta-DC}Non-Gaussian parts of the transition rates at $\epsilon=\omega_0$ for infinitesimal $\eta$ (black solid curve) and with pure dc voltage $N=0$ at low temperature. Curves with different finite broadenings are also shown for comparison.}
\end{figure}
%
%
\begin{figure}[htp]
	\centering
	\includegraphics[scale=0.9]{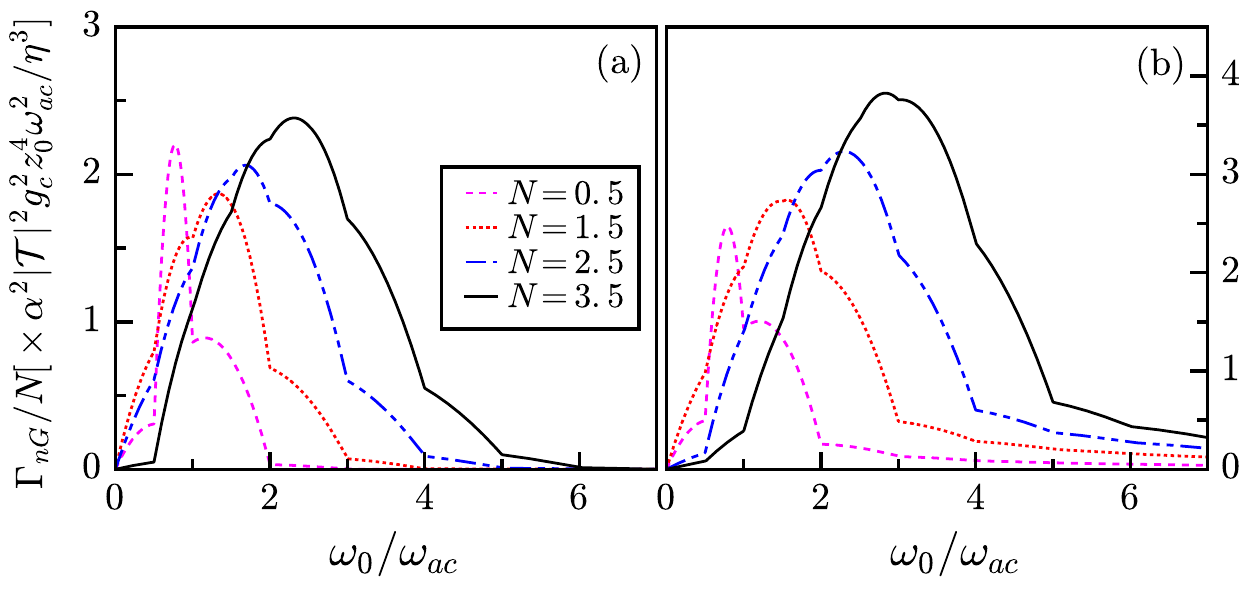}
	\caption{\label{fig:LE-N-nG-zero-eta}  Non-Gaussian parts of the transition rates for infinitesimal $\eta$ with $e\bar{V}=\omega_{ac}$ at low temperature for different $N$, for (a) harmonic drive and (b) square-wave drive. }
\end{figure}
%
%
\subsection{Infinitesimal broadening}
Here we consider the limit where $\eta\rightarrow0$. In this limit, the emission rates at $\epsilon=\omega_0$ are most relevant. Thus we can take the limit $\epsilon\rightarrow\omega_0$ for the Gaussian rate. Now equation~(\ref{eq:GammaG}) becomes $\Gamma_G(\omega_0)=2\pi\alpha^2|\cT|^2(z_0^2/\eta^2)S_{tot}(\omega_0)$, which is just proportional to the non-symmetrized noise with a scaled prefactor. While for the non-Gaussian rate, it becomes more delicate. To proceed, we take the following procedures. First, we take the limit $\epsilon\rightarrow\omega_0$, which will cancel the third term in equation~(\ref{eq:GammaNG}). Then we take the limit $\eta\rightarrow0$, and the impedance inside the integrand become $|z_\omega|^2=(\pi z_0^2\omega_0^2/2\eta)[\delta(\omega+\omega_0)+\delta(\omega-\omega_0)]$ and $\mathrm{Re}(z_\omega)=(\pi/2)z_0\omega_0[\delta(\omega+\omega_0)+\delta(\omega-\omega_0)]$. Now the non-Gaussian rate becomes $\Gamma_{nG}(\omega_0)=(\pi^2/4\eta^3)\alpha^2|\cT|^2g_c^2z_0^4[\cW_{ac}(\omega_0)-W(\omega_0)][-2\cW_{ac}(\omega_0)-\cW_{ac}(0)+3\cW_{ac}(2\omega_0)]$. First we consider the case with pure dc-drive as shown in figure~\ref{fig:LE-nG-zero-eta-DC}. We can see the non-Gaussian rate at $\epsilon=\omega_0$ for infinitesimal $\eta$ only gives a nonzero value when $e\bar{V}/2<\omega_0<e\bar{V}$. Surprisingly, no overbias light emission occurs for infinitesimal $\eta$. The inclusion of finite broadening smears out this feature. Then we consider the case with the effect of ac drive as shown in figure~\ref{fig:LE-N-nG-zero-eta}. Unlike the pure dc-driven case, the inclusion of the ac drive broadens the range of the non-Gaussian rate. Extra kinks appear at $\omega_0=n\omega_{ac}$, which is due to ac-excited-photon-assisted tunneling as discussed before. From figure~\ref{fig:LE-N-nG-zero-eta}, it also can be seen that for the square-wave drive the non-Gaussian rate has a long tail compared to the harmonic drive case, which is due to the fact that the higher order harmonics of $|\cg_{k}|^2$ for square-wave drive decay much slower than those for harmonic drive as shown in figure~\ref{fig:gk-distribution}(c) and (d).
%
%
\section{Summary}
We have investigated the quantum current noise of a coherent conductor in the presence of arbitrarily-shaped time-dependent drives. We have shown that ac-noises possess distinctly different features compared to the dc-case. For instance, an ac drive will smear out the cutoff of the noise at $\Omega=e\bar{V}$. Extra kinks at $\Omega=e\bar{V}+n\omega_{ac}$ appear, which can be explained by photon-assisted tunneling. We have generalized the excess quantum noise to the finite-frequency case and found the minima in the excess noise are shifted to $N=n+\Omega/\omega_{ac}$ and extra kinks arise at $N=n-\Omega/\omega_{ac}$.

Furthermore, we have extended the DCB theory to describe light emission in an ac-driven tunnel junction. The results for the single-electron light emission captured by the Gaussian rates can be well explained by photon-assisted tunneling. Because the Gaussian rates are proportional to the non-symmetrized quantum noise, extra kinks at $\epsilon=e\bar{V} +n\omega_{ac}$ appear, which can be decomposed into processes in which one electron tunnels inelastically through the barrier and contributes to the light emission by absorbing $n$ energy quanta with probabilities $|\cg_n|^2$. The Gaussian rates also depend on the shapes of the ac-drives. For a sharp Lorentzian drive, the Gaussian rates exhibits a  resonant peak even when the applied drive is small. For the non-Gaussian rates, the results become more distinguishable for different shapes of the voltage drives. In the end, we have considered the case of infinitesimal broadening. Surprisingly, there is no overbias light emission even in the pure dc case.

%

%
\section{Acknowledgments}
We gratefully acknowledge the support from DFG through SFB 767 and the Zukunftskolleg of the University of Konstanz.
%




%
\section*{References}
\bibliographystyle{unsrt}
\bibliography{references}
%

\end{document}